\documentclass{article}
\usepackage{spconf,amsmath,graphicx,amssymb,url,subcaption,comment,color}

\usepackage{algorithm}
\usepackage{algpseudocode}
\usepackage{comment}

\algnewcommand{\algorithmicand}{\textbf{ and }}
\algnewcommand{\algorithmicor}{\textbf{ or }}
\algnewcommand{\OR}{\algorithmicor}
\algnewcommand{\AND}{\algorithmicand}

\def\x{{\mathbf x}}
\def\h{{\mathbf h}}
\def\y{{\mathbf y}}

\title{Run-and-back stitch search: novel block synchronous decoding for streaming encoder-decoder ASR}

\name{Emiru Tsunoo$^{\star}$ Chaitanya Narisetty$^{\dagger}$ Michael Hentschel$^{\star}$ Yosuke Kashiwagi$^{\star}$ Shinji Watanabe$^{\dagger}$}
\address{$^{\star}$ Sony Group Corporation, Japan \\
      $^{\dagger}$ Carnegie Mellon University, USA}
\begin{document}
\ninept
\maketitle
\begin{abstract}
%A streaming style inference of encoder--decoder automatic speech recognition (ASR) systems has garnered significant attention.
A streaming style inference of encoder--decoder automatic speech recognition (ASR) systems is important for reducing latency, which is essential for interactive use cases.
To this end, we propose a novel blockwise synchronous decoding algorithm with a hybrid approach that combines endpoint prediction and endpoint post-determination.
In the endpoint prediction, we compute the expectation of the number of tokens that are yet to be emitted in the encoder features of the current blocks using the CTC posterior.
Based on the expectation value, the decoder predicts the endpoint to realize continuous block synchronization, as a {\it running stitch}.  
Meanwhile, endpoint post-determination probabilistically detects backward jump of the source--target attention, which is caused by the misprediction of endpoints.
Then it resumes decoding by discarding those hypotheses, as {\it back stitch}.
We combine these methods into a hybrid approach, namely {\it run-and-back stitch search}, which reduces the computational cost and latency.
%Evaluations of the Librispeech English, AISHELL-1 Mandarin, and CSJ Japanese tasks show the efficiency of the proposed decoding algorithm while maintaining recognition accuracy.
Evaluations of various ASR tasks show the efficiency of our proposed decoding algorithm, which achieves a latency reduction, for instance in the Librispeech test set from 1487 ms to 821 ms at the 90th percentile, while maintaining a high recognition accuracy.
%The proposed method archives up to 54.6\% of reduction on response time, without significant degradation of word error rates.
\end{abstract}
\section{Introduction}
\label{sec:intro}
In recent years, end-to-end automatic speech recognition (ASR) has garnered significant attention.
For interactive use cases in particular, streaming style inference is essential; thus, several approaches have been discovered for both the encoder--decoder (Enc--Dec) \cite{chorowski15, chan16,watanabe17} and transducer models \cite{graves13rnnt,rao17}.
Blockwise processing can be easily introduced to the encoders of both models \cite{miao2020,moritz20,povey18,tsunoo19,shi2021emformer}.
Although transducers are efficient for streaming ASR owing to frame-synchronous decoding, %huge computational resources are required for training.
%In addition, 
they are less accurate than Enc--Dec \cite{li2020comparison} and Enc--Dec can be used additionally to achieve higher performance \cite{sainath2019two}.
%In addition, Enc--Dec architecture is easier to integrate with back-end applications such as machine translation \cite{berard2016listen}.
However, during blockwise Enc--Dec inference, it is still challenging for the decoder to know when to stop decoding, that is, endpoints with limited encoder features in the currently given blocks.

Several studies on streaming Enc--Dec ASR have introduced additional training or modules to predict endpoints.
By predicting endpoints, the decoder realizes continuous block synchronization with the encoder, which we refer to as the {\it running stitch} approach.
Monotonic chunkwise attention (MoChA) \cite{chiu2017monotonic} is a popular approach for achieving online processing \cite{miao2020, kim19, inaguma20}. 
However, it makes training complicated and sometimes degrades accuracy \cite{kim19,inaguma20}.
A triggered attention mechanism \cite{moritz20} and its modification with a scout network \cite{wang20v_interspeech} % and continuous integrate-and-Fire (CIF) \cite{dong20} are
are further methods for predicting the endpoints, but they also require additional modules in the network and training criteria.

As an alternative approach, block synchronization with the encoder can also be established by examining the tendency of hypotheses that exceed the endpoint \cite{tian20, tsunoo2021slt}.
If an exceeded hypothesis is detected, %the decoder starts over from the previous hypotheses with extended encoder features, to which we refer as a {\it back stitch} approach.
the decoder turns back to the previous hypotheses and restarts the beam search with the subsequent encoder features, which we refer to as a {\it back stitch} approach.
Tian et. al. trained the model to emit $\langle$eos$\rangle$ tokens when the decoder exceeded endpoints \cite{tian20}.
In our previous work, we revealed that detecting only $\langle$eos$\rangle$ is insufficient for some ASR tasks, and showed that it is also necessary to detect repeated tokens during decoding \cite{tsunoo2021slt}.
This method achieved better performance than most {\it running-stitch} approaches, and it did not require any additional training.
However, hypotheses might be discarded during decoding and latency can become large, particularly for long utterances because there are real token repetitions.

In this study, we propose a novel blockwise synchronous decoding algorithm that does not require any additional training, with a hybrid approach combining endpoint prediction and endpoint post-determination, namely the {\it run-and-back stitch} (RABS) search.
To realize endpoint prediction, % without additional training,
the CTC posterior and source--target (ST) attention of the decoder within the CTC/attention framework \cite{watanabe17} are used.
We compute the expectation of the number of tokens to be emitted in the current encoder output blocks to predict the endpoint.
We recover a possible misprediction of this endpoint using the post-determination approach.
To achieve this, we explicitly compute the probability of a backward jump in the ST attention to improve on our previous repeated phrase detection.
By combining the {\it running stitch} and {\it backward stitch} in the hybrid RABS search, we can reduce the computational cost and latency.
Experiments on the Librispeech English, AISHELL-1 Mandarin, and CSJ Japanese tasks demonstrate that the proposed RABS search achieves a reduction in latency without significant degradation in word error rates (WERs).
We particularly reduce the 90th percentile latency, in the Librispeech test set for instance, from 1487 ms to 821 ms.

\begin{figure*}[t]
%\vspace{-3.5cm}
  \hspace{0.2cm}
  \begin{minipage}[t]{0.30\linewidth}
  \centering
  %\hspace{1.5cm}
  \includegraphics[width=1.1\columnwidth]{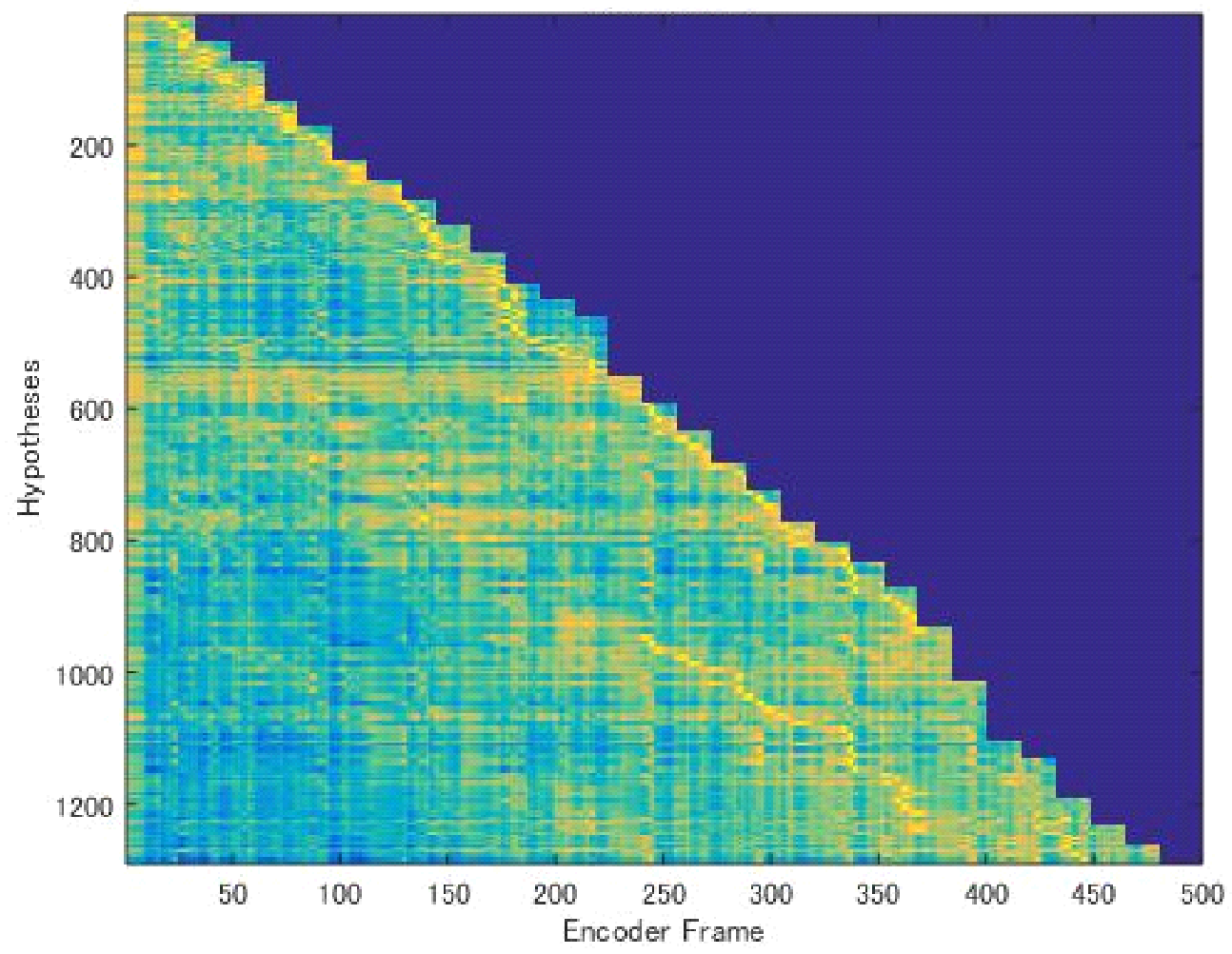}
  \vspace{-0.3cm}
  \subcaption{ST attention of {\it running stitch} search.  The decoder repeats a sequence after approximately hypothesis 900.}
  \label{fig:jump}
  \end{minipage}
  \hspace{0.3cm}
  \begin{minipage}[t]{0.30\linewidth}
  \centering
  \includegraphics[width=1.1\columnwidth]{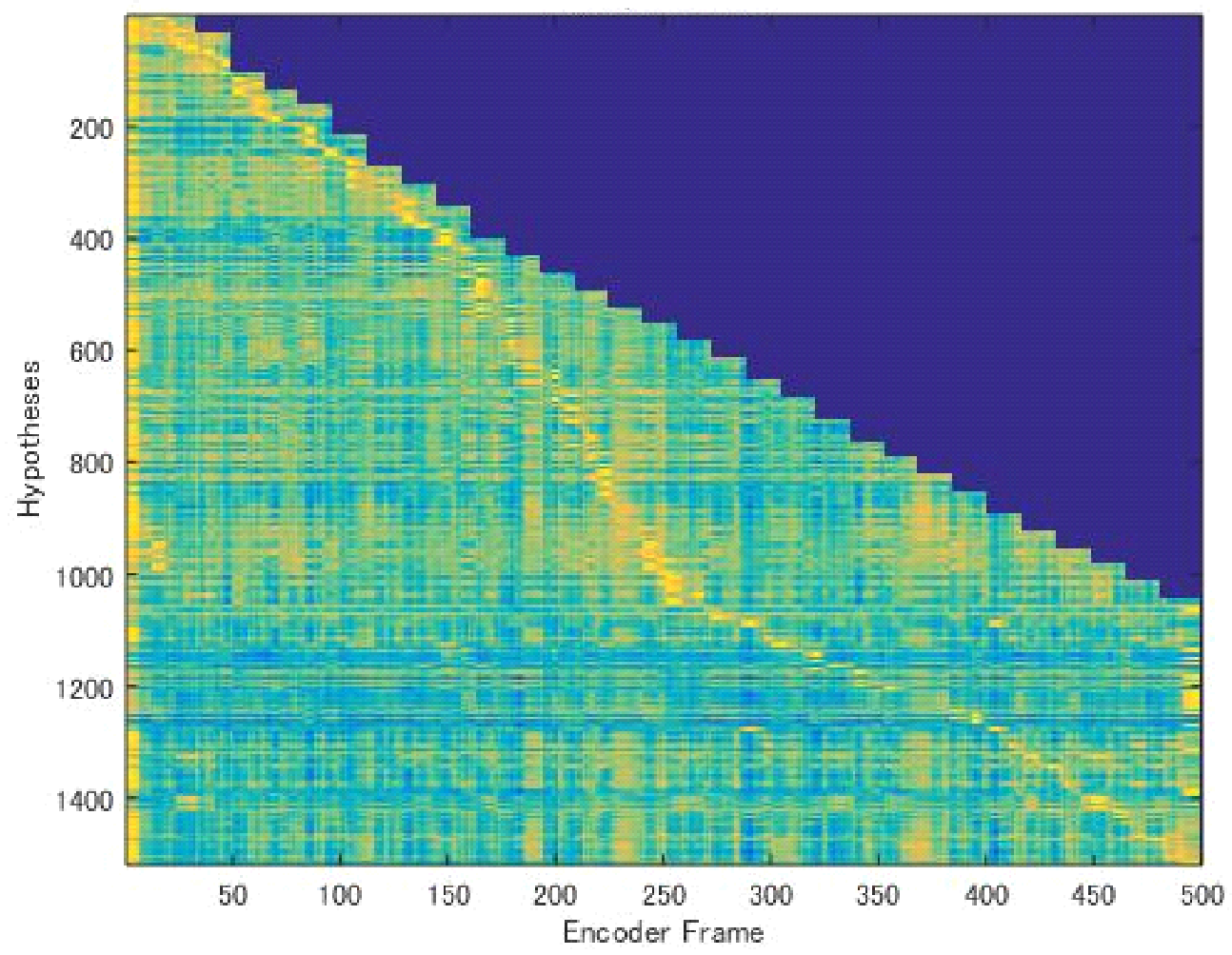}
  \vspace{-0.3cm}
  \subcaption{ST attention of block synchronous beam search.  The decoder struggles after approximately hypothesis 400, whereas the encoder proceeds forward.}
  \label{fig:struggle}
  \end{minipage}
  \hspace{0.3cm}
  \begin{minipage}[t]{0.30\linewidth}
  \centering
  %\hspace{1.5cm}
  \includegraphics[width=1.1\columnwidth]{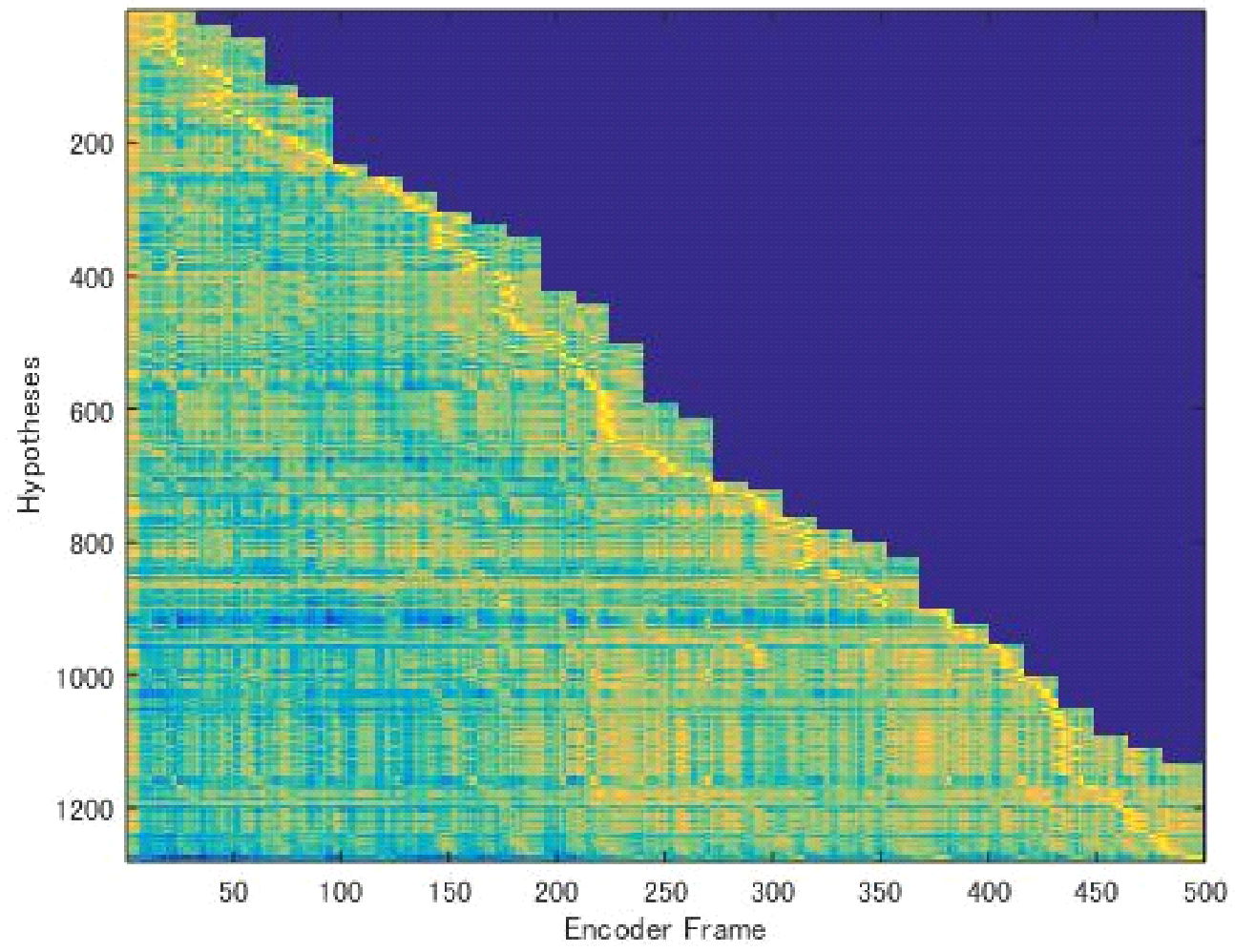}
  \vspace{-0.3cm}
  \subcaption{ST attention of {\it run-and-back stitch} search.  The attention is synchronously aligned to the blockwise encoder.}
  \label{fig:correct}
  \end{minipage}
  \caption{Examples of ST attentions.}
  \vspace{-0.3cm}
\end{figure*}

\section{Streaming encoder-decoder ASR}
\label{sec:ASR}
To realize a streaming Enc--Dec ASR system, both the encoder and decoder must be processed online synchronously.
A simple way to process the encoder online is through blockwise computation, as in \cite{miao2020,moritz20,povey18}.
However, the global channel, speaker, and linguistic context are also important for local phoneme classification.
Therefore, a context inheritance mechanism for block processing was proposed in \cite{tsunoo19,shi2021emformer} by introducing an additional context embedding vector in the encoder.
% The context embedding vector is computed in each layer of each block and handed over to the upper layer of the following block.
Thus, the encoder sequentially computes encoded features $\h_{1:T_b}$ from the currently given $b$ block input $\x_{1:T_b}$. % as % and its context embeddings $\mathbf{c}_{1:b}$, as
\begin{comment}
\vspace{-0.2cm}
\begin{align}
    % \h_{1:b} = \mathrm{Enc}(\x_{1:T_b},\mathbf{c}_{1:b}).
    \h_{1:T_b} = \mathrm{Enc}(\x_{1:T_b})
\end{align}
\end{comment}
%Thus, with the number of left context $N_{\mathrm{l}}$ and that of future context $N_{\mathrm{r}}$, the block input sequence $\mathbf{u}_b=(x_{T{b-1}-N{\mathrm{l}}},\dots,x_{Tb+N{\mathrm{r}}})$ for block $b$ with the context embedding vector is encoded into features $\h_b=(h_{T_{b-1}+1},\dots,h_{T_b})$, where $T_b$ is a index for the last frame of block $b$.

The synchronous decoding in this work bases on the Transformer architecture proposed in \cite{tsunoo2021slt}.
The decoder predicts the probability of the subsequent character from the previous output characters $\y_{0:i-1}$ and the current encoder output blocks $\h_{1:T_b}$, as
\vspace{-0.2cm}
\begin{align}
    p(y_i|\y_{0:i-1},\h_{1:T_b}) = \mathrm{Dec}(\y_{0:i-1},\h_{1:T_b}).
\end{align}
Self-attention in the decoder attends to the output history, $\y_{0:i-1}$, and the subsequent ST attention is directed to the encoder output sequence, $\h_{1:T_b}$.
In addition to the Enc--Dec Transformer, a linear layer is added to the encoder to project $\h_{1:T_b}$ onto the token probability for CTC, which is jointly trained as in  \cite{watanabe17}.

\section{RABS decoding search}% for streaming Enc--Dec ASR}
\label{sec:decoding}

\begin{figure}[t]
%\vspace{-3.5cm}
  \centering
  %\hspace{1.5cm}
  \includegraphics[width=1.0\columnwidth]{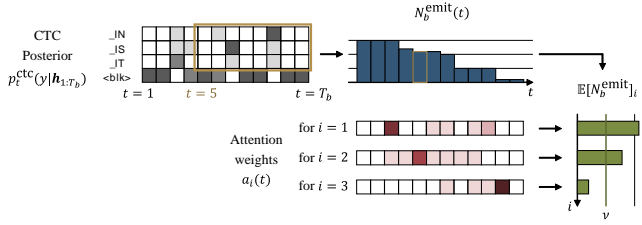}
  \vspace{-0.3cm}
  \caption{Endpoint prediction by expecting the number of tokens in the encoded features using a CTC posterior and a ST attention.}
  \label{fig:ctc}
  \vspace{-0.3cm}
\end{figure}

\subsection{Running stitch approach: endpoint prediction with CTC posterior}
\label{ssec:running}
Endpoint prediction realizes continuous and efficient block synchronization.
It would be a trivial problem to predict endpoints if the system knows the number of tokens to be emitted from the given limited partial encoder features; the decoder should only stop after a specific number of tokens are decoded.
However, because the length of the input sequence and that of the output sequence differ and depend on token granularity, the number of tokens in the encoded features is unknown.
Although most studies introduce an additional endpoint predictor into the ASR model \cite{miao2020,moritz20, chiu2017monotonic}%, dong20}
, we predict endpoints without additional training or modules, within the standard CTC/attention architecture.
%The expectation of the number of token to be emitted is computed with a CTC posterior and a ST attention weight, as shown in Fig.~\ref{fig:ctc}.
We compute an expectation value of the number of tokens following the currently attending time frame in the currently given encoded blocks.
The expectation value is calculated by combining a CTC posterior and an ST attention.

CTC computes a frame-wise token emission posterior, including a blank label.
Let $p_t^{\mathrm{ctc}}(y|\h_{1:T_b})$ be a CTC posterior of frame $t$ given encoded block $\h_{1:T_b}$, as shown in Fig.~\ref{fig:ctc}.
In CTC, the same consecutive tokens are merged, and only the tokens transiting from the other tokens or the blank token are counted.
Because the probability of the previous token not being $y$ is $1-p_{t-1}^{\mathrm{ctc}}(y|\h_{1:T_b})$, the probability of emitting token $y$ in frame $t$ is described as follows:
\begin{align}
    e_y(t) = \left(1-p_{t-1}^{\mathrm{ctc}}(y|\h_{1:T_b})\right) \cdot p_{t}^{\mathrm{ctc}}(y|\h_{1:T_b})
    \label{eq:emit}
\end{align}
Let $N^{\mathrm{emit}}_{b}(t)$ be the remaining number of all the tokens except for the blank to be emitted from the encoded features $\h_{1:T_b}$ after frame $t$.
The number can be expressed as an accumulation of the token emission probability (\ref{eq:emit}) as follows:
\vspace{-0.2cm}
\begin{align}
    N^{\mathrm{emit}}_{b}(t) = \sum_{\tau=1}^{T_b-t}\sum_{y\neq \langle\mathrm{blk}\rangle} e_y(t+\tau)
\end{align}
In the case of Fig.~\ref{fig:ctc}, to compute the number of tokens after $t=5$, all the posterior except for the blank in the yellow box is accumulated for $N^{\mathrm{emit}}_{b}(t=5)$.
$N^{\mathrm{emit}}_{b}(t)$ is shown as the blue graph in Fig~\ref{fig:ctc}.

We use ST attention to know the time frame currently attended to.
An averaged attention weight for the current hypothesis can be computed from ST attention as follows:
\vspace{-0.2cm}
\begin{align}
    a_{i}(t) = \sum_{m=1}^{M}\mathrm{softmax}\left(\frac{q_{m,i-1}k_{m,t}}{\sqrt{d}}\right),
\end{align}
where $q_{m,i-1}$ denotes the query value of the last output in the hypothesis, $y_{i-1}$, $k_{m,t}$ is the key value of encoded frame $t$ and of multi-head $m$ of $M$ heads, and $d$ is the dimensionality of both vectors.  
In Transformer, the ST attention of each head is not always monotonic or aligned, because of the multi-head and residual connections.
However, empirically, we found that the ST attention of the last decoder layer tends to be monotonic. %; thus we use only the ones of the last layer of the decoder.
%Finally, the expectation of the number of token left can be computed as follows.
Thus, we compute the expectation of the remaining number of tokens after currently attending time frame as follows:
\vspace{-0.2cm}
\begin{align}
    \mathbb{E}[N^{\mathrm{emit}}_{b}]_i = \sum_{t=1}^{T_b}a_{i}(t) N^{\mathrm{emit}}_{b}(t) \label{eq:expectation}
\end{align}
%In the case of $i=3$ in Fig.~\ref{fig:ctc}, the ST attention $a_3(t)$ already attends the end of the current encoded features, and the expectation value becomes $\mathbb{E}[N^{\mathrm{emit}}_{b}]_3=0.2$.
%Thus, if the expectation $\mathbb{E}[N^{\mathrm{emit}}_{b}]_i$ is less than threshold $\nu$, it is regarded as an endpoint and the decoder stops until the encoder outputs the next block $\h_{b+1}$.
As shown as the green graph in Fig.~\ref{fig:ctc}, the expectation values $\mathbb{E}[N^{\mathrm{emit}}_{b}]_i$ %, are calculated with $N^{\mathrm{emit}}_{b}$ and $a_{i}(t)$, which 
decrease step by step.
If %the expectation 
$\mathbb{E}[N^{\mathrm{emit}}_{b}]_i$ is less than the predefined threshold $\nu$, we assume that current decoding step $i$ reaches an endpoint of the current $b$ blocks, and the decoder stops until the encoder outputs the next block $\h_{T_b+1:T_{b+1}}$.

\subsection{Back stitch approach: endpoint post-determination}
\label{ssec:postdet}
\subsubsection{Block synchronous beam search}
\label{sssec:bsdec}
CTC and ST attention are not explicitly trained to be aligned; thus, the aforementioned endpoint prediction can cause some errors.
Fig.~\ref{fig:jump} shows an example of ST attention of endpoint prediction search.
The horizontal axis denotes the encoder frame and the vertical axis indicates the hypothesis number.
In the earlier hypotheses, the decoder uses only the limited encoded features, e.g., $\h_{1:T_5}=\h_{1:96}$ for hypothesis 200.
In this example, the ST attention jumps back and the decoder repeats a sequence after approximately hypothesis 900.
The endpoint post-determination can recover such error.
For this purpose, block synchronous (BS) beam search \cite{tsunoo2021slt} can be applied, in which the repeated tokens are considered as well as $\langle$eos$\rangle$ to evaluate the excess of the endpoints.
%The tokens in the hypotheses are regarded as signs of exceeding the endpoint, and thus, we discard the hypotheses containing such tokens and resume decoding after the next block is encoded.
Detecting such tokens in the hypotheses is regarded as signs of exceeding the endpoint, and thus, we discard the hypotheses containing such tokens and resume decoding after the next block is encoded.

Fig.~\ref{fig:struggle} shows an example of ST attention in the BS search, which shows another problem caused by repeated token detection.
The decoder struggles after hypothesis 400, whereas the encoder proceeds forward.
After the encoder reaches the end of the utterance at approximately hypothesis 1000, the decoder consumes all the remaining encoded features, which adversely impacts the latency.
This particularly occurs in long utterances because there are several real token repetitions, which make the decoder discard hypotheses falsely and be left behind by the encoder.
In this study, we improve the BS search by explicitly evaluating ST attention to find the back-jump phenomenon in Fig.~\ref{fig:jump}.

\subsubsection{Attention back jump detection}
\label{sssec:attbs}
When the back-jump phenomenon occurs, the peak of the current ST attention focuses on encoder features prior to the last decoded hypothesis.
Thus, the probability of the attention back jump, $p^{\mathrm{jump}}_{i,b}$, is calculated by accumulating all the ST attention that is concentrated behind the previous attention, as follows.
\vspace{-0.2cm}
\begin{align}
    p^{\mathrm{jump}}_{i,b} = \sum_{t=1}^{T_b} a_{i}(t) \left(\sum_{\tau=1}^{T_b-t} a_{i-1}(t+\tau)\right)
\end{align}
If the back jump probability is greater than a threshold, that is, $p^{\mathrm{jump}}_{i,b}>\upsilon$, the current hypotheses ending with $y_i$ are regarded as a repeated sequence and discarded.
The decoder stops until the next block $\h_{T_b+1:T_{b+1}}$ is encoded.

\subsection{Hybrid RABS beam search}
\label{ssec:rabs}
To realize efficient yet accurate ASR decoding, we propose to combine the endpoint predictor using CTC posterior and the endpoint post-determination with attention back jump detection.
The proposed hybrid beam search algorithm, namely RABS search, is summarized in Algorithm \ref{alg:decode}.
In every beam search step, first, $\langle$eos$\rangle$ and back jump probability of ST attention are evaluated in the current hypotheses (line 6).
If $\langle$eos$\rangle$ is found in the hypotheses or $p^{\mathrm{jump}}_{i,b}>\upsilon$, the decoder is considered to exceed the endpoint and stops decoding until the next block $\h_{T_b+1:T_{b+1}}$ to be encoded.
Subsequently, endpoint prediction is performed by evaluating the expectation of the remaining number of tokens to be emitted, as in (\ref{eq:expectation}), with a CTC posterior and a ST attention (line 14).
An example of ST attention in the proposed RABS search is shown in Fig.~\ref{fig:correct} in which the attention of the decoder proceeds synchronously with the encoder.

\begin{algorithm}[t]
\caption{Hybrid run-and-back stitch beam search}   
\label{alg:decode}                  
\footnotesize
\begin{algorithmic}[1]     
\Require encoder feature blocks $\h$, total block number $B$, beam width $K$, max output length $I_{\mathrm{max}}$
\Ensure $\hat{\Omega}$: complete hypotheses
\State \textbf{Initialize:} $y_0\gets\langle \mathrm{sos}\rangle$, $\Omega_{0} \gets \{y_{0}\}$, $b\gets 1$, $i\gets1$
\While{$b < B$}
\State {$\mathrm{NextBlock}\gets false$}
%\FOR{$i\gets I_{b-1}+1$ to $I_{b}$ unless $\mathrm{NextBlock}$}
\State $\Omega_{i} \gets \mathrm{Search}_{K}(\Omega_{i-1},\h_{1:T_b})$
\For{{$\y_{0:i} \in \Omega_{i}$}}
\If{$yi=\langle \mathrm{eos}\rangle$ \OR $p^{\mathrm{jump}}_{i,b}>\upsilon$} \Comment{\textcolor{red}{\bf back-stitch search}}
\State $\mathrm{NextBlock}\gets true$
\EndIf
\EndFor
\If{$\mathrm{NextBlock}$}
\State $b \gets b + 1$ \Comment{discard current hypotheses and wait for the next block}
% \State \textbf{continue}
% \EndIf
\Else
\If{$\mathbb{E}[N^{\mathrm{emit}}_{b}]_i<\nu$} 
\Comment{\textcolor{red}{\bf running-stitch search}}
\State $b \gets b + 1$ \Comment{wait for the next block}
\EndIf
\State $i \gets i + 1$
\EndIf
\EndWhile
\While {$i<I_{\mathrm{max}}$ unless $\mathrm{EndingCriterion}(\Omega_{i-1})$} \Comment{ordinary decoding follows to obtain $\hat{\Omega}$ after $b=B$}
\State $\Omega_{i} \gets \mathrm{Search}_{K}(\Omega_{i-1},\h_{1:T_B})$  
\For{$\y_{0:i}\in\Omega_{i}$}
\If{$y_{i}=\langle \mathrm{eos} \rangle$}
\State $\hat{\Omega} \gets \hat{\Omega} \cup \y_{0:i}$ 
\EndIf
\EndFor
\EndWhile
\State \Return $\hat{\Omega}$
\end{algorithmic}
%\vspace{-2pt}

\end{algorithm}

\begin{table*}[t]
  \caption{WERs and computation efficiency in the LibriSpeech task. %(L) indicates the equivalent model size as the large model, and (LL) is larger than the large model. 
  ($\ast$: There is no description of the beam size in the literature. $\dagger$: The EPs were evaluated on a different dataset and by simulation without considering computation time.) }
  \label{tab:librispeech}
  %\vspace{1mm}
  \vspace{-0.3cm}
  \centering
%  \begin{tabular}{l|cc|cc}
  \scalebox{0.9}{
  \begin{tabular}{l|cc|cc|c|cc|c}
    \hline
     & \multicolumn{2}{c|}{Beam 30} & \multicolumn{2}{c|}{Beam 10} & RTF & \multicolumn{2}{c|}{Latency} & Avg. last \\
     & test-clean & test-other & test-clean & test-other && EP50 & EP90 & steps\\
    \hline\hline
    %\multicolumn{2}{l}{Batch processing}  \\
    %\hline
    %ContextNet \cite{han2020} (SOTA) & 2.1 & 4.6 & 1.9 & 4.1 \\  % large LM 07-11
    %Transformer \cite{karita19} (ours) & 2.5 & 6.3 & 2.8 & 6.4 \\  % large LM 07-11
    % \ \ \ {\it w/ Transforemr LM} & 2.4 & 5.9 & 2.7 & 6.1 \\  
    
    % CBP-ENC + Batch Dec. \cite{tsunoo19} & 2.7 & 7.2 & 2.9 & 7.3 \\ % Transformer LM
    
    %\multicolumn{5}{l}{Streaming processing with contextual blockwise encoder \cite{tsunoo19}} \\
    % CIF + Chunk-hopping \cite{dong20}  &--- &--- & 3.3 & 9.6 &---&--- &--- &--- \\
    Triggered Attention \cite{moritz20}  & {2.8} & {7.3} & ---&--- &---& ---&--- & ---\\ % Niko
    HS-DACS \cite{li2021head}  & 2.7& 6.6& --- &--- &---&--- &--- &--- \\ 
    Scout Network \cite{wang20v_interspeech} & ---&--- & {2.7} & {6.4}  &---& ---&--- & ---\\ % 
    Emformer Transducer (pretrained in hybrid ASR) \cite{shi2021emformer} & --- &--- & ${\bf 2.4}^\ast$ & ${\bf 6.1}^\ast$ &---&--- &--- &--- \\ 
    FastEmit Conformer Transducer \cite{yu2021fastemit} &--- &--- & $3.5^\ast$ & $9.1^\ast$ &--- & $290^\dagger$ ms & $660^\dagger$ ms & ---\\
    \hline 
    \multicolumn{8}{l}{Regular model (6-layer decoder with $d=256$ and $M=4$)} \\
    \hline
    % CTC \cite{tsunoo19} & 3.3 & 9.1 & && & & &N/A \\ %
    BS-Dec \cite{tsunoo2021slt} & {\bf 2.7} & {\bf 7.1} &{\bf 3.0} &{\bf 7.7}& 0.25 & 552 ms & 1487 ms & 7.95\\ % espnet1 
    % BS-Dec \cite{tsunoo2021slt} & {\bf 2.9} & {\bf 7.3} &{\bf 3.0} &{\bf 7.7}& 0.25 & 552 ms & 1487 ms & 8.28\\ % espnet2 
    Running-stitch search (Sec.~\ref{ssec:running}) & 3.0 & 7.7 & 3.3 &8.5& {\bf 0.24} & {\bf 349 ms} & {\bf 508 ms} & {\bf 3.09} \\
    Back-stitch search (Sec.~\ref{ssec:postdet}) & {\bf 2.7} & {\bf 7.1} & {\bf 3.0}&{\bf 7.7}& 0.25 & 497 ms & 919 ms & 5.22 \\ % espnet1
    % Back-stitch search (Sec.~\ref{ssec:postdet}) & {\bf 2.9} & {\bf 7.3} & {\bf 3.0}&{\bf 7.7}& 0.25 & 497 ms & 919 ms & 5.08 \\ % espnet2
    RABS search (proposed) & 2.8 & 7.2 &{\bf 3.0} & 7.8& {\bf 0.24} & {491 ms} & {821 ms} & 4.71 \\ % espnet1
    % RABS search (proposed) & {\bf 2.9} & 7.4 &{\bf 3.0} & 7.8& {\bf 0.24} & {491 ms} & {821 ms} & 5.24 \\ % espnet2
    \hline
    \multicolumn{8}{l}{Small model (2-layer decoder with $d=256$ and $M=4$)} \\
    \hline
    BS-Dec \cite{tsunoo2021slt} & {\bf 2.9} & {\bf 7.5} & {\bf 3.4}&8.4& 0.15 & 341 ms & 857 ms & 8.29\\ % LM Large redone on 10 July 2020
    % Back-stitch search (proposed) & {3.0} & {7.4} & 3.1&8.2& 0.16 & 357 ms & 521 ms & XXX \\
    RABS search (proposed) &3.0 & 7.6& {\bf 3.4} &{\bf 8.3}& {\bf 0.14} & {\bf 246 ms} & {\bf 354 ms} & {\bf 3.25}\\ % {\bf 272 ms} & {\bf 418 ms} & rs with max\\
    % \ \ \ {\it w/ Transformer LM} & {\bf 2.3} & {\bf 6.5} & {\bf 2.6} & {\bf 6.7} \\ % Transformer LM redone on 22 June 2020 Transformer LM
    \hline
    \multicolumn{8}{l}{Large model (6-layer decoder with $d=512$ and $M=8$)} \\
    \hline
    HuBERT BS-Dec + Transformer LM & {\bf 2.2} & {\bf 4.3} & {\bf 2.2} &{\bf 4.4}&---  &--- &---& 12.28 \\
    HuBERT RABS search + Transformer LM (proposed) &{\bf 2.2} &{\bf 4.3} & 2.3 & 4.5& --- &--- &---&{\bf 1.52}\\
    \hline
  \end{tabular}
  }
  \vspace{-0.2cm}
\end{table*}

\section{Experiments}
\subsection{Experimental Setup}
\label{ssec:setup}
We conducted experiments using the English LibriSpeech dataset \cite{panayotov15}, AISHELL-1 \cite{aishell17} Mandarin tasks, and the Japanese CSJ dataset \cite{csj}.
The input acoustic features were 80-dimensional filter bank features and the pitch.
Regular, small, and large models were trained using multitask learning with CTC loss as in \cite{watanabe17} with a weight of 0.3.
%, using ESPNet \cite{watanabeespnet}.
% A linear layer was added to the encoder to project $\h$ onto the token probability for CTC. 
We used the Adam optimizer and Noam learning rate decay, % as in \cite{vaswani17}, 
and applied SpecAugment \cite{park19}.
%Training was performed using ESPNet % \footnote{The training and inference implementations are publicly available at \url{https://github.com/espnet/espnet}.} 
%\cite{watanabeespnet} .

We adopted contextual block processing with a Transformer encoder, following \cite{tsunoo2021slt}. % \cite{tsunoo19}.
% The block-wise encoder computation was performed following \cite{tsunoo2021slt}.
For the regular and small models, we trained a 12-layer encoder with 2048 units, $d=256$ of the hidden dimension size, and $M=4$ multihead attention.
The large model was trained with $d=512$ and $M=8$ and used HuBERT \cite{hsu2021hubert} features \footnote{We used causal feaures for decoding.} pretrained on Libri-light \cite{kahn2020libri}.

The decoder had six layers with 2048 units for the regular and large models, and two layers for the small model.
The parameters for the ordinary (batch) decoder were directly used in the proposed hybrid RABS algorithm of the decoder.
We set the parameters for the RABS search as $\nu=1.0$ and $\upsilon=0.5$.

The real-time factor (RTF) and latency were measured with our implementation of the proposed search algorithm in C++, using a subset of each task with a beam size of 10.
We only evaluated regular and small models because we did not implement the large model with HuBERT in C++.
%We used Intel Math Kernel Library to perform matrix operations with CPUs.
% A caching technique was applied as in \cite{tsunoo2021slt}.
We adopted EP latency, following \cite{yu2021fastemit}, which is the time required to emit $\langle \mathrm{eos}\rangle$ token after the end of each utterance.
The 50th and 90th percentile latencies are shown. 
The RTF and latency were measured with an 8 core 3.60 GHz Intel i9-9900K processor.

\begin{table}[t]
  \caption{CERs and latency in the AISHELL-1 task}
  \label{tab:aishell}
  %\vspace{1mm}
  \vspace{-0.3cm}
  \centering
%  \begin{tabular}{l|cc|cc}
  \scalebox{0.9}{
  \begin{tabular}{l|cccc}
    \hline
     & Dev  & Test & EP50 & EP90\\
    \hline\hline
    RNN-T \cite{tian19} & 10.1 & 11.8 &---&---\\
    Sync-Transformer (6-layer) \cite{tian20} & {7.9} & 8.9&---&--- \\
    % CBP-ENC + MoChA Dec. \cite{tsunoo2019towards}  & 9.7 & 9.7 &---&--- \\ %TODO: redo evaluation
    HS-DACS \cite{li2021head} & 6.2 & 6.8 &--- &--- \\
    BS-Dec \cite{tsunoo2021slt} & {\bf 5.8} & {\bf 6.4} &439 ms & 570 ms\\
    %RABS search (proposed) & {\bf 5.8} &{\bf 6.4} & {\bf 357 ms} & {\bf 549 ms} \\
    RABS search (proposed) & {\bf 5.8} &{\bf 6.4} & {\bf 326 ms} & {\bf 471 ms} \\
    \hline
  \end{tabular}
  }
  %\vspace{-0.4cm}
\end{table}

\begin{table}[t]
  \caption{CERs and latency in the CSJ task}
  \label{tab:csj}
  %\vspace{1mm}
  \vspace{-0.3cm}
  %\centering
  \hspace{-0.3cm}
%  \begin{tabular}{l|cc|cc}
  \scalebox{0.9}{
  \begin{tabular}{l|ccccc}
    \hline
     & eval 1 & eval 2 & eval 3 &EP50 & EP90\\
    \hline\hline
   % CBP-ENC + MoChA Dec. \cite{tsunoo2019towards} &  &  &  \\
    % CBP-ENC + CTC \cite{tsunoo19} & 6.2 & 4.5 & 5.2 & &\\
    % BS-Dec \cite{tsunoo2021slt} & {5.3} & {4.1} & {4.5} &1027 ms &2924 ms\\
    % BS-Dec \cite{tsunoo2021slt} & {5.5} & {4.3} & {4.8} &1027 ms &2924 ms\\ espnet1
    BS-Dec \cite{tsunoo2021slt} & {\bf 5.5} & {\bf 4.2} & {\bf 4.7} &978 ms &2616 ms\\ % espnet2
    % RABS search (proposed) & 5.6 &4.3 &4.8 &536 ms & 995 ms \\ espnet1
    RABS search (proposed) & 5.6 & {\bf 4.2} &{\bf 4.7} & {\bf 582 ms} & {\bf 1079 ms} \\ % \\&536 ms & 995 ms \\ % espnet2

    \hline
  \end{tabular}
  }
  %\vspace{-0.4cm}
\end{table}

\subsection{Librispeech English results}
For LibriSpeech, we adopted the byte-pair encoding subword tokenization \cite{sennrich16}, which has 5000 token classes.
A language model (LM) of a four-layer LSTM with 2048 units was fused with a weight of 0.6 for the regular and small models.
For the large model, a 16-layer Transformer LM was fused as in \cite{tsunoo2021slt}.
CTC weight was set as 0.4.
We compare WERs with a beam size of 30 and that of 10.
The running-stitch approach described in Sec.~\ref{ssec:running}, the back-stitch approach described in Sec.~\ref{ssec:postdet}, and their combination, that is, RABS search (Sec.~\ref{ssec:rabs}), were compared with our previous BS search \cite{tsunoo2021slt}.

% Lastly, the number of decoding step after the encoder reached the utterance end, i.e., the number of steps after line 20 in Algorithm~\ref{alg:decode}, was evaluated.
We also evaluated the average number of last decoding steps after the encoder reached the utterance end, that is, the number of steps after line 19 in Algorithm~\ref{alg:decode}.
We desire as few decoding steps as possible, but if the encoder and decoder are not sufficiently synchronous, as shown in Fig.~\ref{fig:struggle}, the value increases.
%We evaluated the average of the number of steps in the aforementioned subset.

%The results are shown in Table~\ref{tab:librispeech}.
The results are listed in Table~\ref{tab:librispeech} along with other streaming approaches with a larger number of parameters \cite{moritz20,shi2021emformer,wang20v_interspeech,li2021head,yu2021fastemit}.
The results of the regular models show that the running-stitch search is efficient as RTF and latency decreased, particularly at EP90.
However, WERs increased owing to the CTC posterior and ST attention misalignment. 
Back-stitch search successfully decreased the average number of last steps, by replacing repeated token detection with the proposed attention back jump probability while maintaining the WERs of BS search.
%We observed further reduction with RABS search, which recovered the error caused by the Running-stitch approach.
With the RABS search, we observed a WER reduction compared with the running-stitch search because it successfully recovered the error by the back-stitch approach.
In addition, the latency of RABS search improved over the back-stitch search, and the EP90 latency decreased compared with BS search, from 1487 ms to 821 ms.
% In addition, we achieved the best latency median (EP50) among the regular models. 
The same tendency can be found in the small model.
In particular, the 10-beam RABS search with the small model was faster and more accurate than the FastEmit transducer \cite{yu2021fastemit}.
% We could confirm that those models were feasible enough for interactive use-cases.
% By comparing WERs with other methods with a larger number of parameters, our proposed RAB search achieved comparable accuracy. % with small latency.
We also confirmed that our proposed approach can be applied to the large model with HuBERT features, which drastically reduced the number of last steps without significant WER degradation.

\subsection{AISHELL-1 Mandarin and CSJ Japanese results}
We used the regular model architectures for AISHELL-1 Mandarin and CSJ Japanese tasks to confirm the effectiveness of our proposed method in other languages, i.e., in various token granularity.
For the Mandarin task, 4231 character classes were used with parameters \{CTC weight, beam width, LM weight\} = \{0.5, 10, 0.7\}.
For CSJ, the dataset had 3260 Japanese character classes,
and parameters were set as \{CTC weight, beam width, LM weight\} = \{0.3, 10, 0.3\}.
We used an external two-layer LSTM LM with 650 units for each tasks.

The results for AISHELL-1 are summarized in Table~\ref{tab:aishell} and for CSJ are in Table~\ref{tab:csj}.
We observed a similar tendency as on Librispeech; the proposed RABS search reduced latency while WERs were maintained. 
In the Mandarin task, we achieved better performance compared to other streaming methods, such as head-synchronous decoding (HS-DACS)~\cite{li2021head}.
We confirmed that our proposed method is consistently effective in various languages.

\section{Conclusion}
We have proposed a novel blockwise synchronous decoding algorithm for Enc--Dec ASR, called RABS search, which is a hybrid approach combining endpoint prediction and endpoint post-determination.
In the endpoint prediction, the expectation of the number of tokens that are yet to be emitted in the encoder features of currently given blocks is calculated. %, with which the synchronous decoding is efficiently carried out.
The endpoint post-determination recovers errors from endpoint misprediction, by copmputing backward jump probability of the ST attention.
%The combined RABS search reduced the computational cost and latency due to the former while maintaining accuracy due to the latter.
The RABS search successfully combined the advantages of both, reducing the computational cost and latency, and maintaining accuracy at the same time.
% Evaluations of various ASR confirmed that the proposed approach can be applied various languages and token generalities.
%Evaluations of various ASR tasks show the efficiency of the RABS search, which achieved a latency reduction without sacrificing accuracy. 
%Evaluations on Librispeech, AISHELL-1, and CSJ tasks showed the effectiveness of our proposed RABS across multiple languages.
%The proposed RABS search achieved latency reduction without significant degradation of WERs.

%Future work includes an improvement on self-attention computation in the decoder, because it is still quadratic and not efficient for long input sequences.
%In future work, we want to focus on an improvement in the self-attention in the decoder because its quadratic complexity makes it currently inefficient for long input sequences.

\bibliographystyle{IEEEbib}
\bibliography{mybib}

\end{document}